\documentclass[varenna]{cimento}
\usepackage{graphicx}  
\title{Exotic atom pairs: Repulsively bound states in an optical lattice}
\author{ J.~Hecker Denschlag}
    \institute{ Institut f\"ur
Experimentalphysik, Universit\"at Innsbruck, 6020 Innsbruck, Austria}
\author{A. J.~Daley} \institute{Institut f\"ur Theoretische Physik, Universit\"at Innsbruck, 6020 Innsbruck,
Austria} \PACSes{\PACSit{03.75.Lm, }\PACSit{42.50.-p}{\ldots}}

\newcommand{\ket}[1]{|#1\rangle}
\newcommand{\beq}{\begin{equation}}
\newcommand{\eeq}{\end{equation}}

\newcommand{\rmi}{{\rm i}}

\begin{document}

\maketitle


\section{Introduction}
Ultracold atoms in 3D optical lattices provide an intriguing environment to
study strongly correlated condensed matter systems and quantum information.
Unique features of these atomic many body systems include the complete control
of system parameters, and -– in particular contrast to solid state physics -–
weak couplings to dissipative environments. This so-called quantum lattice gas
\cite{hubbardtoolbox,Blo05} is described by a Bose or Fermi Hubbard
Hamiltonian. The high control over the atoms opens the possibility to engineer
a wide class of interesting many body quantum states. Seminal experiments have
already demonstrated the superfluid-Mott insulator transition \cite{Gre02}, the
realisation of 1D quantum liquids with atomic gases \cite{Par04,Sto04} (see
also \cite{Kin04,Lab04}), and a Bose spin glass \cite{Fal06}. Here we review
another recent experiment \cite{Win2006} where we have observed a novel kind of
bound state of two atoms which is based on repulsive interactions between the
particles. These repulsively bound pairs exhibit long lifetimes, even under
conditions when they collide with one another. Stable repulsively bound objects
should be viewed as a general phenomenon and their existence will be ubiquitous
in cold atoms lattice physics. Although the experiment described here is based
on bosonic $^{87}$Rb atoms, other composites with fermions \cite{Hof02} or
Bose-Fermi mixtures \cite{Lew04} should exist in an analogous manner.
Furthermore, repulsively bound objects could also be formed with more than two
particles.

In the following we will first explain the theoretical background of
repulsively atom pairs. Afterwards we will present the  experiments
which demonstrate several key properties of the pairs. Finally we
give a short discussion of how these repulsively bound pairs relate
to bound states in some other physical systems.

\section{Repulsively bound pairs}

\begin{figure}[htbp]
\begin{center}
\includegraphics[width=0.6\textwidth]{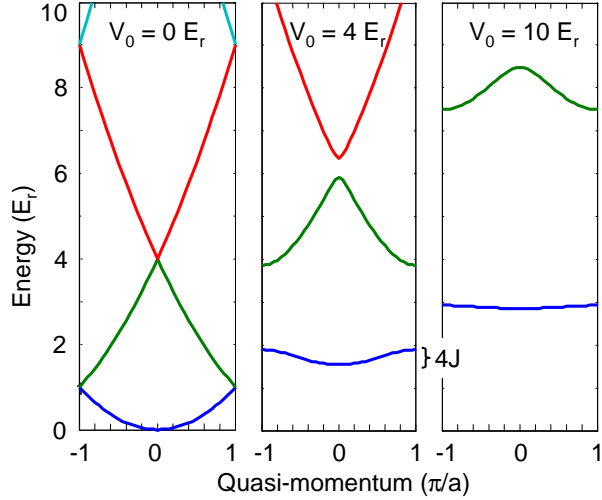}
\caption{Energy spectrum for a single particle in a 1D lattice for three
different potential depths $V_0$. The band width of the lowest Bloch band is
given by $4J$ where $J$ is the hopping energy. $a$ is the lattice period and
$E_r = \pi^2 \hbar^2/ 2 m a^2$ denotes the recoil energy.
}\label{Fig:bandstructure}
\end{center}
\end{figure}

Optical lattices are generated by pairs of counterpropagating laser beams,
where the resulting standing wave intensity pattern forms a periodic array of
microtraps for the cold atoms, with period given by half the wavelength of the
light, $\lambda /2$. This periodicity of the potential gives rise to a
bandstructure for the atom dynamics with Bloch bands separated by band gaps,
which can be controlled via the laser parameters and configuration, as shown in
Fig.~\ref{Fig:bandstructure}. The dynamics of an atomic Bose-Einstein
condensate loaded into the lowest band of a sufficiently deep optical lattice
\cite{hubbardtoolbox,Blo05} is well described by a single band Bose Hubbard
model \cite{Fis89} with Hamiltonian
\begin{equation}
\hat{H}=-J\sum_{\langle ij\rangle }{\hat{b}}_{i}^{\dag }{\hat{b}}_{j}+\frac{U%
}{2}\sum_{j}{\hat{b}}_{j}^{\dag }{\hat{b}}_{j}\left( {\hat{b}}_{j}^{\dag }{%
\hat{b}}_{j}-1\right) +\sum_{i}\epsilon _{i}{\hat{b}}_{j}^{\dag }{\hat{b}}%
_{j}.  \label{BH}
\end{equation}%
Here ${\hat{b}}_{i}$ (${\hat{b}}_{i}^{\dag }$) are destruction (creation)
operators for the bosonic atoms at site $i$. $J/\hbar$ and $U$ denote
respectively the tunnelling rate of atoms between neighbouring sites, and the
collisional energy shift from interactions between atoms on the same site. The
resulting width of the Bloch band is $4J$, and this single band model is valid
because the kinetic energy and interaction energy in this system are much
smaller than the separation of the Bloch bands $\omega$.

\begin{figure}[htbp]
\begin{center}
\includegraphics[width=\textwidth]{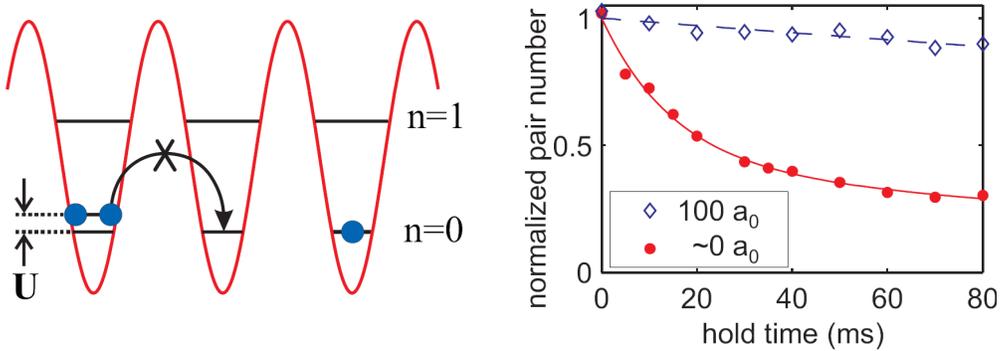}
\caption{{\bf left} A state with two atoms located on the same site
of an optical lattice has an energy offset $\approx U$
  with respect to states where the atoms are separated. Breaking up of
the pair is suppressed due to the lattice band structure and energy
conservation, so that the pair remains bound as a composite object,
which can tunnel through the lattice. In the figure, $n=0$ denotes
the lowest Bloch band and $n=1$ the first excited band. {\bf right}
Breaking up of atom pairs in a shallow 3D optical lattice. [The
potential depth is $ (10\pm 0.5)\,E_r$ in one direction and $(35\pm
1.5)\,E_r$  in the perpendicular directions.] Shown is the remaining
fraction of pairs for a scattering length of 100\,a$_{0}$ (open
diamonds) and a scattering length of about $(0\pm 10)$\,a$_{0}$
(filled circles) as a function of the hold time. The atom pairs
quickly break up within a few ms if they do not interact, but stay
together for a long time in case of repulsive interaction.
 The lines are fit curves of an exponential (dashed
line) and the sum of two exponentials (solid line). }
\label{Fig:pairbreak}
\end{center}
\end{figure}

The Bose-Hubbard Hamiltonian (\ref{BH}) predicts the existence of
stable repulsively bound atom pairs. These are most intuitively
understood in the limit of strong repulsive interaction $U>>J$
(where $U>0$ but this energy is still smaller than the separation to
the first excited Bloch band, $U\ll \omega$). If a state is prepared
with two atoms occupying a single site, $|2_{i}\rangle \equiv
({\hat{b}}_{i}^{\dag }{}^{2}|$vac$\rangle)/\sqrt{2} $ , then it will
have a potential energy offset $\approx U$ with respect to states
where the atoms are separated [see Fig. \ref{Fig:pairbreak}(left)].
This state will be unable to decay by converting the potential
energy into kinetic energy, as the Bloch band provides a maximum
kinetic energy for two atoms both at the edge of the Brillouin zone
given by $8J<<U$. Instead, the atoms will remain together, and
tunnel through the lattice as a bound composite object -- a
repulsively bound pair.

We can observe this nature of repulsive binding in the experiment
[see Fig.~\ref{Fig:pairbreak}(right)]. After production of the atom
pairs, as is discussed in detail in section
\ref{sec:exprealisation}, we allow the atoms to tunnel through the
lattice along one dimension. If the on-site interaction of the atoms
is tuned to zero with the help of a Feshbach resonance, the pairs
break up within a few ms, corresponding to the tunneling timescale.
However, if the effective interaction between the atoms is
repulsive, we observe a remarkably long lifetime of $t = 700$\,ms
(determined by an exponential fit). This lifetime is mainly limited
by inelastic scattering of lattice photons.

\section{Analytical solution of two particle problem in an optical lattice}
\label{section:analytical}

\subsection{General Discussion}
Our understanding of these stable pairs can be made more formal by an exact
solution of the two particle Lippmann-Schwinger scattering equation on the
lattice corresponding to the Bose-Hubbard Hamiltonian Eq.~(\ref{BH}).

Denoting the primitive lattice vectors in each of the $d$ dimensions by
$\mathbf{e}_{i}$, we can write the position of the two atoms by
$\mathbf{x}=\sum_{i=1}^dx_{i} \mathbf{e}_{i}$ and $\mathbf{y}=\sum_{i=1}^dy_{i}
\mathbf{e}_{i}$, where
$x_{i}, y_i$ are integers, and we can write the two atom wave function in the form $%
\Psi (\mathbf{x},\mathbf{y})$. The related Schr\"odinger equation from the
Bose-Hubbard model [Eq.~(\ref{BH})] with homogeneous background, $\epsilon_i=0$
then takes the form
\begin{equation}
\left[ -J \left(\tilde{\Delta}_{{\bf x}} + \tilde{\Delta}_{{\bf y}} \right) + U
\delta_{{\bf x},{\bf y}}\right] \Psi({\bf x},{\bf y}) = E \: \Psi({\bf x},{\bf
y}), \label{schroedinger}
\end{equation}
where the operator
\begin{equation}
\tilde{\Delta}_{{\bf x}} \Psi({\bf x},{\bf y})\!=\! \sum^{d}_{i=1}\left[
\Psi({\bf x\!+\!e}_{i},{\bf y}) \!+\! \Psi({\bf x\!-\!e}_{i},{\bf y}) - 2
\Psi({\bf x},{\bf y})\right]
\end{equation}
denotes a discrete lattice Laplacian on a cubic lattice. Note that
in order to express this in terms of the discrete lattice Laplacian
we have added $4d J \Psi({\bf x},{\bf y})$ to each side of the
Schr\"odinger equation. This effectively changes the zero of energy,
so that $E\rightarrow E+4Jd$. We then introduce relative coordinates
${\bf r} = {\bf x} - {\bf y}$ existing on the same lattice structure
as the co-ordinate ${\bf x}$, and center of mass coordinates ${\bf
R} = ({\bf x} + {\bf y})/2$, existing on a lattice with the same
symmetry as the original lattice but smaller lattice constant $a/2$.
We then separate the wavefunction using the ansatz
\begin{equation}
\Psi({\bf x}, {\bf y}) = \exp(i {\bf K} {\bf R}) \psi_{\bf K}({\bf r}),
\end{equation}
with $\bf K$ the centre of mass quasi-momentum. This allows us to
reduce the Schr\"odinger equation to a single particle problem in
the relative coordinate,
\begin{equation}
  \left[ -2 J \tilde{\Delta}^{\bf K}_{\bf r}  + E_{\bf K}
  + U \delta_{{\bf r},0}\right]
\psi_{\bf K}({\bf r}) = E \psi_{\bf K}({\bf r}) \label{singleparticleSE}
\end{equation}
where $E_{\bf K}= 4 J\sum_{i=1}^d \left[1\!-\!\cos ({\bf K}{\bf e}_{i}/2)
\right]$ is the kinetic energy of the center of mass motion, and where the
discrete lattice Laplacian for a square lattice is now given by
\begin{equation}
\tilde{\Delta}^{{\bf K}}_{{\bf r}} \Psi({\bf r})\!=\! \sum^{d}_{i=1}
\cos\left({\bf K}{\bf e}_{i}/2\right)\left[ \Psi({\bf r\!+\!e}_{i})
\!+\! \Psi({\bf r\!-\!e}_{i}) - 2 \Psi({\bf r})\right].
\end{equation}

The solutions of this Schr\"odinger equation can be found using the Greens
function of the non-interacting problem with $U=0$, which is defined by
\begin{equation}
\left[E - H_{0}\right] G_{\bf K}(E,{\bf r}) =  \delta_{{\bf r},0},
\label{greensfunction}
\end{equation}
with $\delta_{{\bf r},0}$ a three-dimensional Kronecker delta, and
$H_{0}= - 2 J \Delta^{\bf K}_{\bf r}$ the Hamiltonian of the
non-interacting system. This equation can be easily solved via
Fourier transformation, $G(E,{\bf r})=[1/(2\pi)^d] \int {\rm d}^dk
\tilde G(E,{\bf k}) \exp(i {\bf k} {\bf r}$), and we obtain the
solution
\begin{equation}
\tilde G_{\bf K}(E,{\bf k}) = \frac{1}{E- \epsilon_{\bf K}({\bf k}) + i \eta},
\end{equation}
where $\epsilon_{\bf K}({\bf k})$ accounts for the dispersion relation of the
non-interacting system,
\begin{equation}
  \epsilon_{\bf K}({\bf k})= 4 J \sum_{i=1}^{d} \cos\frac{K_{i} a}{2}\left[1-\cos(k_{i} a)\right].
\end{equation}
The solutions of Eq.~(\ref{singleparticleSE}) can be divided into two classes:
scattering states, and bound (localised) states. We we will first analyze the
scattering states.

\subsection{Scattering States}

\begin{figure}[tbp]
\begin{center}
\includegraphics[width=0.9\textwidth]{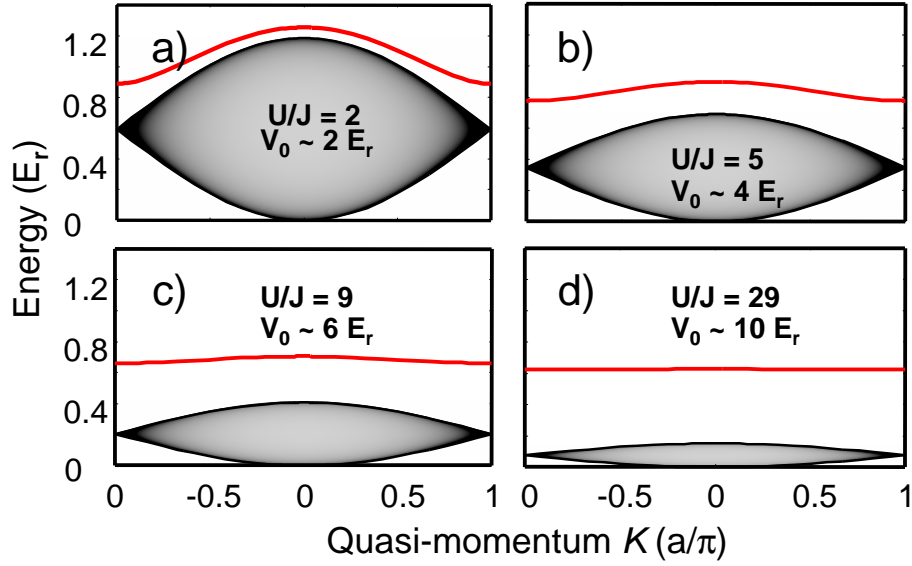}
\caption{Two particle energy spectrum in a 1D lattice for four
different potential depths $V_0$ as a function of centre of mass
quasi-momentum $K$. The Bloch band for repulsively bound pairs is
located above the continuum of unbound scattering states. The grey
level for the shading of the continuum is proportional to the
density of states.}\label{Fig:twoparticleband}
\end{center}
\end{figure}

Similarly to scattering problems involving particles in free space, the
scattering states of particles on the lattice with energy $E$ obey the
Lippmann-Schwinger equation
\begin{equation}
 \psi_{E}({\bf r}) = \psi_{E}^{0}({\bf r}) + \sum_{{\bf r'}} G_{{\bf K}}(E, {\bf r-r'})
V({\bf r'}) \psi_{E}({\bf r'})
\end{equation}
with $\psi_{E}^{0}= \exp( i {\bf k} {\bf r})$  an eigenstate of $H_{0}$ with
energy $E= \epsilon_{\bf K}({\bf k})$. In the present situation with a short
range potential $V({\bf r})= U \delta_{{\bf r},0}$, the Lippmann-Schwinger
equation can be solved via a resummation of the Born expansion and we obtain
\begin{equation}
 \psi_{E}({\bf r}) = \exp( i {\bf k} {\bf r}) - 8 \pi J f_E({\bf K}) G_{{\bf K}}(E, {\bf r}) \label{scatteringstates}
\end{equation}
with scattering amplitude
\begin{equation}
f_E({\bf K})= -\frac{1}{4\pi}\frac{U/(2J)  }{1- G_{\bf K}(E,0) U},
\end{equation}
where the total energy is $E= \epsilon_{{\bf k},{\bf K}} + E_{{\bf K}}$,
$\epsilon_{{\bf k},{\bf K}}= 4 J \sum_{i=1} \cos({\bf K}{\bf e}_{i}
/2)\left[1-\cos({\bf k}{\bf e}_{i})\right]$,
and
\begin{equation}
G_{\bf K}(E,0)=  \frac{4\pi}{2J}\int \frac{{\rm d}{\bf k}}{(2\pi)^d} \frac{1}{
E/(2J) - 2 \sum_{i=1}^d \cos\frac{K_{i}a}{2} \left(1-\cos k_{i} a \right)}.
\end{equation}
The scattering states $\psi_{E}({\bf r})$ correspond to two free atoms moving
on the lattice and undergoing scattering processes. The corresponding energies
appear as a continuum in Fig.~\ref{Fig:twoparticleband}. In order to make a
connection to the scattering length in free space, we can consider the limit of
small momenta of the incoming plane wave, i.e., ${\bf k} \rightarrow 0$, ${\bf
K} \rightarrow 0$ and $E\rightarrow 0$. Then the solution
(\ref{scatteringstates}) reduces in the limit ${\bf r} \rightarrow \infty$ to
\begin{equation}
 \psi_{E}({\bf r}) \sim \psi_{E}^{0}({\bf r})  + f({\bf k},{\bf k'})  \frac{\exp(i k r)}{r}
 \label{asymptoticSS}
\end{equation}
with the scattering amplitude
\begin{equation}
f({\bf k}, {\bf k'})  = - a_{s}=  -\frac{1}{4 \pi}\frac{U/(2 J)}{1-\alpha U/(2
J)} ,
\end{equation}
equivalent to the $s$-wave scattering length $a_{s}$, while  the constant
$\alpha=\lim_{E\rightarrow 0} G(E,0)$ is $\alpha\approx -0.25$ \cite{alpharef}.

\subsection{Bound States}

Note that the scattering amplitude in Eq.~(\ref{scatteringstates}) contains a
pole, associated with a bound state. We now focus on these bound states in the
regime, $U>0$, which will correspond to a repulsively bound pair. First we note
that we can write Eq.~(\ref{greensfunction}) in the form
\begin{equation}
\left[E - H_{0}\right] G_{\bf K}(E,{\bf r}) =  \frac{1}{G_{\bf
K}(E,0)}\delta_{{\bf r},0}  G_{\bf K}(E,{\bf r}).
\end{equation}
\begin{figure}[tbp]
\begin{center}
\includegraphics[width=\textwidth]{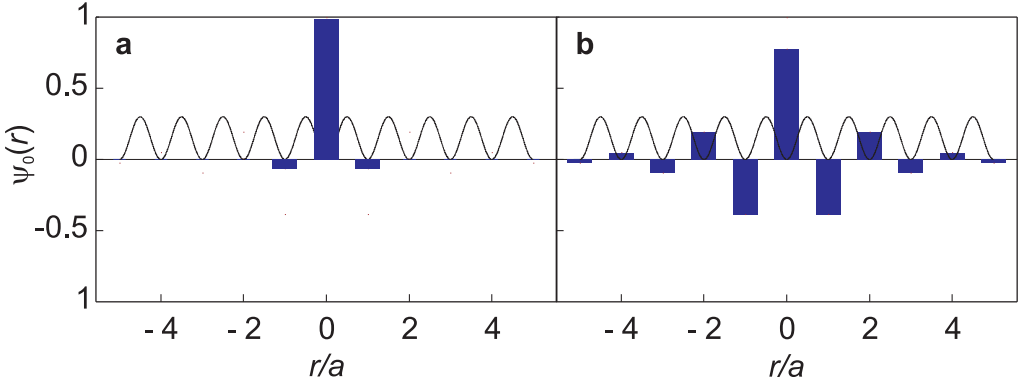}
\caption{ Wavefunctions $\psi_{\bf K}({\bf r})$ showing the
amplitude for various site separations for repulsively bound pairs
($a_s = 100 a_0$) in 1D with $K=0$.  $r$ denotes the separation
between the two atoms. (a) $U/J=30$ $(V_0 = 10 E_r)$ and (b) $U/J=3$
$(V_0 = 3 E_r)$.} \label{Fig:wavefunctions}
\end{center}
\end{figure}

As a consequence, the function $\psi^{\rm BS}_{\bf K}({\bf
r})=G_{\bf K}(E,{\bf r})$ is a solution of the Schr\"odinger
equation if the self-consistency relation is satisfied
\begin{equation}
U =\frac{1}{ G_{\bf K}(E,0)}, \label{selfconsistency}
\end{equation}
which determines the energy $E_{\rm BS}$ of the bound state
$\psi^{\rm BS}_{\bf K}$. The resulting bound state wavefunction,
$\psi^{\rm BS}_{\bf K}({\bf r})$ falls off exponentially for large
$r$, and describes a bound two particle state travelling with center
of mass momenta ${\bf K}$ through the lattice. The momentum
distribution of this bound state is then given by $G_{\bf K}(E_{\rm
BS},{\bf k})$
\begin{equation}
 \psi^{\rm BS}_{\bf K}({\bf k}) = \frac{1}{E_{\rm BS}- \epsilon_{\bf K}({\bf k}) }.
\end{equation}
Note, that this wave function is not normalized.

In three dimensions, Eq.~(\ref{selfconsistency}) only has a solution
for interaction strengths above a critical value, $U> U_{c}= - 2 J /
G(0,0)\approx 8 J$, and thus we require $U>U_c$ for the formation of
the bound two-particle state. The wavefunction $\psi_{\bf K}({\bf
r})$ is then square-integrable, as shown in
Fig.\ref{Fig:wavefunctions}. For a deep lattice, i.e. $U/J \gg 1$,
bound pairs essentially consist of two atoms occupying the same
site, whereas for small $U/J$, the pair is delocalized over several
lattice sites. A main feature of the repulsive pair wavefunction is
its oscillating character: the wavefunction amplitude alternates
sign from one site to the next, as shown in
Fig.~\ref{Fig:wavefunctions}. In quasimomentum space this
corresponds to a wavefunction which is peaked at the edges of the
first Brillouin zone, which is discussed in detail in
section \ref{sec:quasimomentumresults}.

When motion is confined to one dimension the bound two particle state exists
for arbitrarily small repulsive interaction $U>0$, in contrast to the
three-dimensional situation. Here the energy of the bound pairs, computed from
Eq.~(\ref{selfconsistency}) is $E_{LS}(K)=2J[\sqrt{4( \cos \frac{Ka}{2})
^{2}+(U/2J) ^2}+2]$, which can be seen plotted in Fig.
\ref{Fig:twoparticleband} as Bloch band of a stable composite object
\emph{above} the continuum of two particle scattering states. The figure shows
how the binding energy (separation of these states from the continuum)
increases as $U/J$ is increased, and how the curvature of the band becomes less
pronounced. In the limit of strong interaction, $U\gg J$ the bound state energy
reduces to $E(K)\sim 4J+U+(4J^2/U) (1+\cos Ka)$, which is consistent with our
expectation of a positive binding energy $U$, and the center of mass energy of
a composite object with an effective tunnelling matrix element $J^{2}/U$.

\section{Numerical approach for repulsively bound pairs}

We want to be able to treat not just a single repulsively bound pair, but a
lattice gas of many interacting repulsively bound pairs. This is important both
in order to properly describe the effects of interactions on experimental
measurements, and to investigate many-body effects on the behaviour of the
pairs. Whilst perturbation theory can used to produce useful analytical models
in some limits \cite{Fle06}, it is possible in one dimension to treat the
system in more general regimes using recently developed numerical techniques.

\subsection{Time-Dependent DMRG}
In one spatial dimension, the system of interacting repulsively bound pairs can
be treated by directly simulating the Bose-Hubbard model time-dependently,
using time-dependent Density Matrix Renormalisation Group (DMRG) methods
\cite{td-dmrg1,td-dmrg2,td-dmrg3,dmrgreview}. These methods allow near-exact
integration of the Schr\"{o}dinger equation for the 1D many body system by an
adaptive decimation of the Hilbert space, provided that the resulting states
can be written efficiently in the form of a truncated matrix product state
(this will be explained in more detail below). This method has been
successfully applied to several lattice and spin models of interest in
condensed matter physics, including systems realisable using cold atoms in
optical lattices
\cite{vex1,vex2,vex3,vex-spectroscopy,vex4,vex5,vex-spectroscopy2}. The
algorithm, both in the form originally proposed by Vidal and similar methods
proposed by Verstrate and Cirac have also been generalised to the treatment of
master equations for dissipative systems and systems at finite temperature
\cite{Vid04,Cir04}.

\subsubsection{Matrix Product States}
DMRG methods are based on a decomposition of the many-body wavefunction into a
matrix product representation \cite{dmrgreview}. This requires that the
original state can be expressed on a Hilbert space that is the product of local
Hilbert spaces. For the Bose-Hubbard model, each local Hilbert space
corresponds to a particular lattice site, and the basis states in the local
Hilbert space correspond to different occupation numbers, running from 0 to
$S-1$. We then write the coefficients of the wavefunction expanded in terms of
local Hilbert spaces of dimension $S$,
\beq
\ket{\Psi}=\sum_{i_1 i_2 \ldots i_M = 1}^S c_{i_1 i_2 \ldots i_M} \ket{i_1}
\otimes \ket{i_2} \otimes \ldots \otimes \ket{i_M},
\eeq
as a matrix product state given by
\beq
c_{i_1 i_2 \ldots i_M}=\sum_{\alpha_1 \dots \alpha_{M-1}}^\chi \Gamma^{[1] \;
i_1}_{\alpha_1}\lambda^{[1]}_{\alpha_1}\Gamma^{[2] \; i_2}_{\alpha_1
\alpha_2}\lambda^{[2]}_{\alpha_2}\Gamma^{[2] \; i_2}_{\alpha_3 \alpha_4} \ldots
\Gamma^{[M] \; i_M}_{\alpha_{M-1}}.
\eeq
The $\Gamma$ and $\lambda$ tensors are chosen so that the tensor
$\lambda^{[l]}_\alpha$ specifies the coefficients of the Schmidt decomposition
\cite{td-dmrg1} for the bipartite splitting of the system at site $l$,
\beq
\ket{\psi}=\sum_{\alpha=1}^{\chi_l} \lambda^{[l]}_\alpha \ket{\phi_\alpha^{[1
\ldots l]}}\ket{\phi_\alpha^{[l+1 \ldots M]}},
\eeq
where $\chi_l$ is the Schmidt rank, and the sum over remaining tensors specify
the Schmidt eigenstates, $\ket{\phi_\alpha^{[1 \ldots l]}}$ and
$\ket{\phi_\alpha^{[l+1 \ldots M]}}$. The key to the truncated matrix product
state representation is that for many states corresponding to a low-energy in
1D systems we find that the Schmidt coefficients $\lambda^{[l]}_\alpha$, when
ordered in decreasing magnitude, decay rapidly as a function of their index
$\alpha$ \cite{td-dmrg1}. Thus we can truncate the representation at relatively
small $\chi$ and still provide an inner product of almost unity with the exact
state of the system $\ket{\Psi}$. In implementations of this method we perform
convergence tests for the state representation, that is, we vary the values of
$\chi$ and $S$ to check that the point at which the representation is being
truncated does not affect the final results. We also make use of an optimised
version of the code in which the Schmidt eigenstates are forced to correspond
to fixed numbers of particles \cite{td-dmrg2,dmrgreview}. This allows us to
make use of the total number conservation in the Hamiltonian to substantially
increase the speed of the code, and also improve the scaling with $\chi$ and
$S$. With this number conserving code we are able to compute results with much
higher values of $\chi$ in a much shorter time than the original algorithm.

\subsubsection{Time Dependence}

Time dependence of these states can be computed for Hamiltonians acting only on
neighbouring lattice sites because when an operator acts on the local Hilbert
state of two neighbouring sites, the representation can be efficiently updated
to the Matrix product state that best approximates the new state of the system.
To do this, the $\Gamma$ tensors corresponding to those two sites must be
updated, a number of operations that scales as $\chi^3 S^3$ for sufficiently
large $\chi$ \cite{td-dmrg1}. In this way, we represent the state on a
systematically truncated Hilbert space, which changes adaptively as we perform
operations on the state. The time evolution operator $\exp(-\rmi \hat{H} t)$,
is then split into a product of operators, each of which acts only on a pair of
neighbouring sites by means of a Suzuki-Trotter decomposition
\cite{trotter1,trotter2}. This is done in small timesteps $\delta t$. Initial
states can also be found using an imaginary time evolution, i.e., the repeated
application of the operator $\exp(-\hat{H} \delta t)$, together with
renormalisation of the state.

\subsection{Numerical Investigation of Repulsively Bound Pairs}

In investigation of repulsively bound pairs using these methods we are able to
use parameters $U$ and $J$ directly corresponding to values of the lattice
depth $V_0$ in the experiments, and are also able to account for the background
trapping potential. We typically study $10-30$ pairs in $60$ lattice sites, and
begin with an initial product state, corresponding to a random distribution of
doubly-occupied and unoccupied lattice sites. We then reduce the values of $U$
and increase the value of $J$, using the same time dependence for the depth of
the lattice $V_0(t)$ as in the experiment. The single particle momentum
distributions can then be calculated efficiently from the matrix product state
representation, and we also average the results over different initial
configurations, to match the averaging over different 1D tubes in the
experiment. We can also perform lattice modulation spectroscopy, computing the
time evolution of the many-body state when the parameters $U$ and $J$ vary as a
function of time, based on the time dependence of the lattice depth $V(t)$ used
in the experiments (these calculations are similar to those in refs.
\cite{vex-spectroscopy,vex-spectroscopy2}).

\section{Experimental realization}
\label{sec:exprealisation}

In this section we describe the experimental steps to produce a pure ensemble
of repulsively bound atom pairs in an optical lattice. We use $^{87}$Rb as the
atomic species for our experiments.

 \subsection{BEC production}
We begin by creating a $^{87}$Rb Bose-Einstein condensate (BEC) of $6\times
10^5$ $^{87}$Rb atoms in spin state $| F = 1, m_F = -1 \rangle$ in a vacuum
apparatus featuring a magnetic transport line \cite{Tha05,Gre01}. This
transport initially transfers laser cooled atoms from the chamber of the
magneto-optical trap (MOT) into a UHV glass cell (pressure $<$ 10$^{-11}$ mbar)
which offers good optical access from all sides. Here the BEC is produced in a
QUIC trap \cite{QUIC,Tha05} with trapping frequencies $\omega_{x,y,z} = 2 \pi
\times (15, 15, 150)$
 Hz at a magnetic bias field of 2 G. Afterwards the QUIC trap is converted into a
Ioffe-type magnetic trap with trap frequencies $\omega_{x,y,z} = 2
\pi  \times (7, 19, 20)$ Hz) by adjusting the currents through the
quadrupole and Ioffe coils and by applying additional magnetic field
gradients. This moves the BEC over a distance of 8 mm into the
center of the QUIC quadrupole coils which are later used to generate
the homogeneous magnetic field for Feshbach ramping.

\subsection{Loading into lattice}

Within 100\,ms the BEC  is adiabatically loaded into the vibrational
ground state of an optical lattice which is 35\,$E_r$ deep
($E_r=2\pi^2\hbar^2/m\lambda^2$, where m is the mass of the atoms).
Our 3D lattice is cubic and consists of three pairs of
retro-reflected intensity-stabilized laser beams, which propagate
orthogonally to each other. They are derived from a frequency-stable
single-mode Ti:Sapphire laser ($<$ 500 kHz linewidth) with a
wavelength of $\lambda$ = 830.44 nm. For this wavelength, the laser
is detuned by about 100 GHz from the closest transition to an
excited molecular level, minimizing light induced losses as a
precondition for long
 lifetimes of pairs and molecules. The laser beams are polarized
 perpendicularly
to each other and their frequencies differ by several tens of MHz to
avoid disturbing interference effects. The waists of all three beams
are about 160 $\mu$m and the maximum obtainable power is about 110mW
per beam. At this stage about 20\% of the condensate atoms are
grouped in pairs of two into the lattice sites. 60\% of the
condensate atoms are found in singly occupied sites, and another
20\% percent of atoms are located in triply and more highly occupied
lattice sites \cite{Tha06}.

  \subsection{Purification scheme}
In order to remove all atoms from those lattice sites that are not
occupied by exactly two atoms we use a purification scheme which
involves an intermediate step in which Feshbach molecules are
produced. For this, we turn off the magnetic trap and flip the spins
of the $^{87}$Rb atoms from their initial state $| F = 1, m_F = -1
\rangle$ to $|F = 1, m_F = +1 \rangle$ by suddenly reversing the
bias magnetic field of a few G. This spin state features a 210\,mG
wide Feshbach resonance at 1007.40\,G \cite{Volz03}. By
adiabatically ramping over this resonance we convert pairs of atoms
in multiply occupied lattice sites into Rb$_2$ Feshbach molecules
with almost unit efficiency \cite{Tha06}. After the Feshbach ramp,
fast inelastic collisions will occur in sites that were initially
occupied with more than two atoms. These exothermic collisions
between either a created molecule and an atom or between two
molecules will remove these particles from the lattice. At this
stage the lattice consists only of sites which are either empty,
filled with a single atom, or filled with a single Feshbach
molecule. A subsequent 3ms long combined radio-frequency (rf) and
optical purification pulse removes all chemically unbound atoms,
thus creating a pure molecular sample of about $2 \times 10^4$
molecules \cite{Tha06}. The microwave drives the transition at a
frequency of 9113\,MHz between levels which correlate with $| F = 1,
m_F = +1 \rangle$ and $|F=2, m_F = +2 \rangle$. The light pulse
drives the closed transition $| F = 2, m_F = +2 \rangle \rightarrow
| F = 3, m_F = +3 \rangle$. The optical transition frequency is
1402\,MHz blue detuned compared to the transition at zero magnetic
field. The light literally blows the atoms out of the lattice, while
the direct effect of the microwave and light field pulse on the
molecules is negligible because the radiation is off resonance.
Finally, sweeping back across the Feshbach resonance we
adiabatically dissociate the dimers and obtain a lattice where $2
\times 10^4$ sites are filled with exactly two atoms. According to
section \ref{section:analytical}, at the deep lattice depth of 35
$E_r$ ($U/J \approx 3700 $), the corresponding two atom wavepacket
matches perfectly with the wavefunction of the repulsively bound
atom pair. By adiabatically  lowering the lattice depth (typically
within a few ms) in a horizontal direction we can then produce 1D
repulsively bound atom pairs states for arbitrary values of $U/J$.

\section{Experiments}
In the following we discuss the properties of the repulsively bound pairs that
were experimentally investigated by measuring their lifetime, quasi-momentum
distribution and binding energies. By varying the effective interaction between
the atoms with the help of the Feshbach resonance we can also create lattice
induced bound atom pairs which are based on attractive interactions.

\subsection{Pair lifetime}
\label{sec:pairlifetime}

We have already seen from Fig. \ref{Fig:pairbreak}(right) that for a
repulsive interaction with $a_s=100\,a_{0}$ the lifetime of the
pairs is remarkably long ($700$\,ms, exponential fit). This lifetime
is mainly limited by inelastic scattering of lattice photons
\cite{Tha06}, and greatly exceeds the calculated time for an atom to
tunnel from one site to the next, $2\pi\hbar/(4J) \sim 4\,$ms. The
lifetime measurements are based on lowering the lattice depth to a
chosen height, and then measuring the number of remaining pairs
after a variable hold time. In order to make this measurement, the
lattice is adiabatically raised again to its full initial depth of
$V_0 = 35\,E_r$. Using the Feshbach resonance atoms in doubly
occupied sites are converted to Feshbach molecules with near unit
efficiency \cite{Tha06}, and another combined rf-light purification
pulse then removes all remaining atoms (which stem from now
dissociated pairs) as in the original preparation step. Afterwards
the molecules are again converted back into atoms, and can then be
detected via conventional absorption imaging.

\subsection{Quasi-momentum distribution}

\label{sec:quasimomentumresults} 
\begin{figure}[tbp]
\begin{center}
\includegraphics[width=0.9\textwidth]{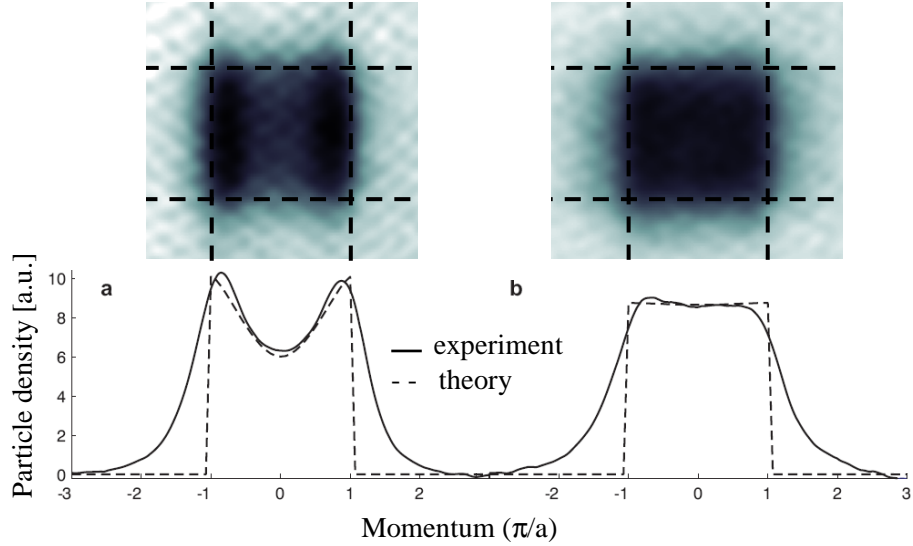}
\caption{Quasi-momentum distributions of atoms in the lattice for
(a) $V_0=5E_R$ and (b) $V_0=20E_R$. {\bf above} Images
 show absorption images of the atomic distribution after release
from the 3D lattice and a subsequent $15\,$ms time of flight. The
horizontal and vertical dashed lines enclose the first Brillouin
zone. {\bf below} Corresponding quasi-momentum distributions  in the
horizontal-direction, after integration over the vertical-direction.
 For comparison numerical simulations (see text) are also shown (dashed lines).
 The density values have been scaled to facilitate comparison between
  experimental and theoretical results. }\label{Fig:momentum}
  \label{Fig:momentumdist}
\end{center}
\end{figure}

\begin{figure}[bp]
\begin{center}
\includegraphics[width=\textwidth]{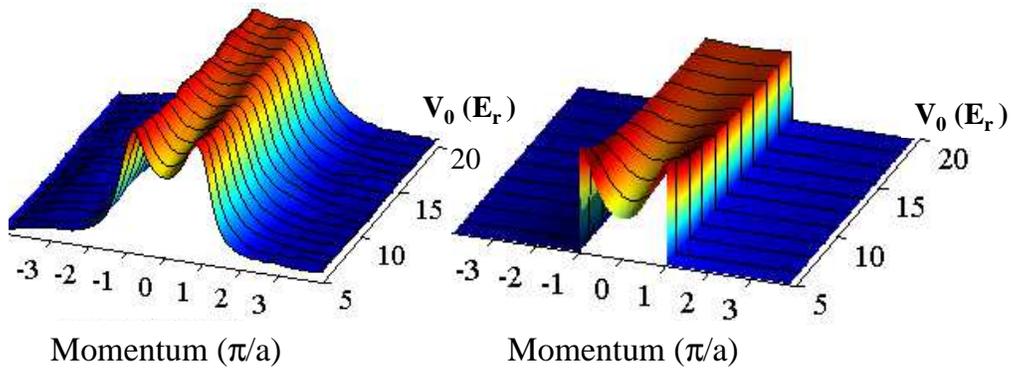}
\caption{Momentum distributions similar as the ones shown in
Fig.\ref{Fig:momentum} are plotted here as a function of lattice
depth $V_0$. Left: experiment. Right: numerical calculation. }
\label{Fig:Qmomentumdist}
\end{center}
\end{figure}

We have experimentally investigated
the quasi-momentum distribution of the pairs in various regimes by
mapping it onto a spatial distribution, which we measured using
standard absorption imaging. For this, we first adiabatically lower
the lattice depth in a horizontal direction at a rate of 1.3 $E_r /
$ms to a pre-chosen height while the lattice depth in the other two
directions are kept high (35 $E_r $). This will prepare repulsively
bound pairs at the chosen lattice depth. We then turn off the
lattice rapidly enough so that the quasi-momentum distribution
cannot change, but slowly with respect to the bandgap, so that
single-particle quasi-momenta are mapped to real momenta
\cite{Gre01,Den02}.  We have typically employed linear ramps with
rates of 0.2 $E_r / \mu$s.  The resulting momentum distribution is
converted to a spatial distribution after $\sim$ 15 ms time of
flight. Fig.~\ref{Fig:momentumdist} shows two measured
quasi-momentum distributions for lattice depths $V_0 = 6 (20)$,
respectively. The top row shows the bare images of the atomic
density taken in the laboratory. Below are the corresponding
quasi-momentum distributions in horizontal direction. It is clearly
visible that for low lattice depths the quasi-momentum distributions
are peaked at the edges of the first Brillouin zone. For deep
lattices, however, the first Brillouin zone is homogenously filled
and the quasi-momentum distribution has a flat top shape. This
latter distribution is reminiscent of the one observed for a
dephased ensemble of ultracold atoms in the lowest band of a lattice
\cite{Gre01}. The agreement between experimental data  and
theoretical calculations is quite good. Note, however, that the
experimental
  distributions appear to extend beyond the first Brillouin zone.
This is an experimental artifact related to repulsion between atoms
during expansion (before imaging) and also relatively long imaging
times (many photons are scattered from each atom, which performs a
random walk). This leads to smearing out of the sharp structure at
the edge of the Brillouin zone.
 Fig.~\ref{Fig:Qmomentumdist}
shows in a more continuous fashion the dependence on lattice depth
$V_0$ of the quasi-momentum distribution for repulsively bound pairs
for both experiment and numerical simulation. As discussed before,
the peak structure is more pronounced for lower values of $V_0$, and
diminishes for larger $V_0$. This characteristic is a clear
signature of the pair wavefunction for repulsively bound pairs.

It is important to note that in all cases here we measure the
distribution of single-atom quasi-momenta in a large sample. That we
still obtain the peaked distribution characteristic of repulsively
bound pairs is non-trivial. In fact, if we just take a single
repulsively bound pair with centre of mass quasi-momentum $K\neq 0$,
its single-atom momentum distribution will not be peaked anymore at
the edges of the first Brillouin zone. The peak will be somewhat
translated towards the center of the first Brillouin zone.
 Fortunately, with increasing $|K|$, the peak in the single-particle
quasi-momentum distribution also becomes less pronounced. As a
result, when we average over a roughly uniform distribution of
centre of mass quasimomenta $K$ for a dilute gas of repulsively
bound pairs, we still observe the pronounced peaks at the edges of
the Brillouin zone. This is confirmed by the numerical simulations
and their good agreement with experiments.

\begin{figure}[p]
\begin{center}
\includegraphics[width=0.8\textwidth]{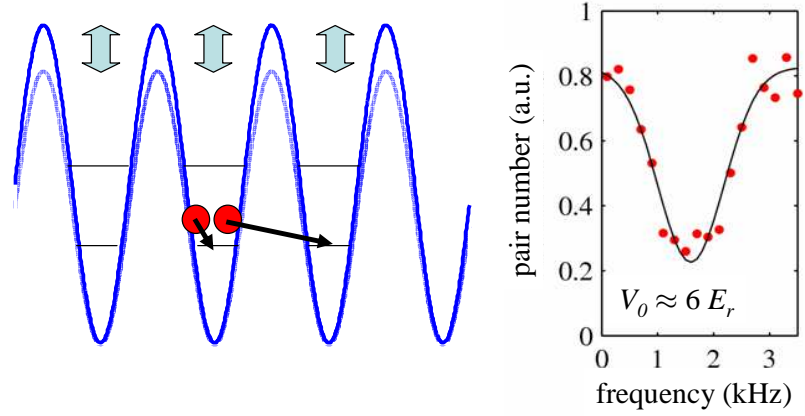}
\caption{Modulation spectroscopy of repulsively bound pairs.
  ({\bf left}) By modulating the optical lattice amplitude
  with the proper frequency, the pair can dump its binding
  energy into the lattice motion and subsequently break up.
  ({\bf right})  Typical resonance dip showing the remaining
 number of atom pairs as a function of the modulation frequency,
  for $V_0\approx 6\,E_r$. The black line is a Gaussian fit,
  a choice which was justified by
numerical calculations.}\label{Fig:modulate}
\end{center}
\end{figure}

\begin{figure}[p]
\begin{center}
\includegraphics[width=0.5\textwidth]{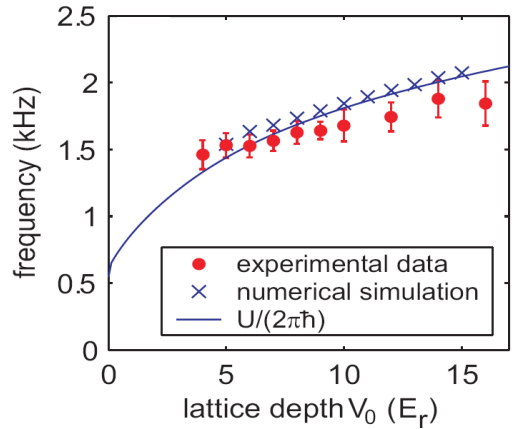}
\caption{Measured resonance frequencies of the modulation
spectroscopy as a function of the lattice depth. The resonance
frequency was determined from resonance curves similar to the one in
 Fig. \ref{Fig:modulate} (right). The experimental points (filled circles)
show good agreement with numerical simulations (crosses) and also
coincide with the onsite collisional energy shift $U$ (line).
Experimental error bars correspond to the 95\% confidence interval
for the Gaussian  fit parameters of the resonance dips.}
\label{Fig:modulate2}
\end{center}
\end{figure}

\subsection{Modulation spectroscopy}

By modulating the depth of the lattice at a chosen frequency we can determine
the binding energy of the pairs (see Fig. \ref{Fig:modulate}, left). For
appropriate modulation frequencies, the pairs can dump their binding energy
into the lattice motion and dissociate. Fig.~\ref{Fig:modulate} (right) shows a
typical resonance curve of the number of remaining pairs as a function of the
modulation frequency. The resonance frequency of about 1.5 kHz (for a lattice
depth of $V_0$ = 6 $E_r$) agrees well with the calculated binding energy of a
pair. The width of the resonance curve can be understood, as the pair will
decay into a continuum of scattering states which has an energy width of up to
$8 J$ [depending on the initial centre of mass quasimomentum $K$ (see Fig.
\ref{Fig:twoparticleband})]. In addition to this width, broadening due to
Fourier limited modulation pulses and inhomogeneity effects in the lattice will
occur.

Modulation spectroscopy measurements were carried out for a variety
of lattice depths (see Fig. \ref{Fig:modulate2}). The resonance
positions are in good agreement with numerical simulations and
essentially coincide with interaction energy, $U$.

\begin{figure}[htbp]
\begin{center}
\includegraphics[width=\textwidth]{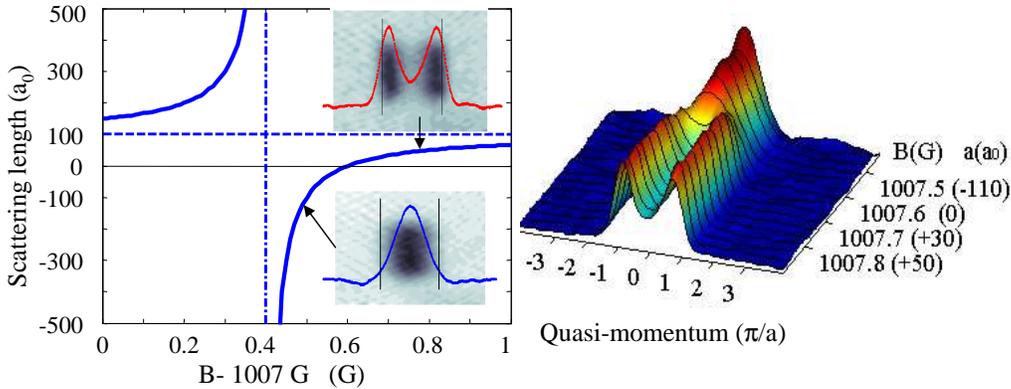}
\caption{From repulsively to attractively bound atom pairs. {\bf
left} With the help of a Feshbach resonance around 1007.4 G, we can
choose the effective interaction of the paired atoms by controlling
the scattering length $a$. $a_0$ is the Bohr radius. The inserted
images show quasi-momentum distributions similar to the ones of Fig.
\ref{Fig:momentum}. For effectively attractive interaction the
quasi-momentum distribution is peaked around 0 momentum. {\bf right}
The momentum distribution for atom pairs as a function of magnetic
field (scattering length.). At zero scattering length the
distribution has a flat top shape. The shown data correspond to
experiments where the lattice depth $V_0$ had been adiabatically
lowered in 1D  below 3$E_r$. }\label{Fig.attractivepair}
\end{center}
\end{figure}

\subsection{Attractively bound pairs}

Making use of the Feshbach resonance at 1007.40 G we can tune the
effective interaction of the atoms within the pair (see Fig.
\ref{Fig.attractivepair}, left). It is then possible to also create
bound atom pairs which are based on attractive interaction. After
initial production of repulsively bound atom pairs in the deep
lattice ($V_0 = 35 E_r$), we applied an appropriate nearly
homogenous magnetic offset field. This tuned the scattering length
of the atomic pair from its  default value of $a_s = 100 a_0$ to
negative scattering length of up to $a_s = -110 a_0$. Afterwards the
optical lattice was lowered as before.  In contrast to repulsively
bound pairs where the momentum distribution is peaked at the edges
of the first Brillouin zone, the momentum distribution for
attractively bound pairs is peaked in the center of the first
Brillouin zone. This goes along with having a bound state with
minimal internal energy. Fig.~\ref{Fig.attractivepair} (right) shows
how the
 quasi-momentum distribution of the pairs changes continuously as
 the scattering length is changed. Interestingly, for
 non-interacting atoms ($a_s = 0$) the distribution again becomes a
 flat top shape.

With respect to stability, we find that lifetimes of bound atom
pairs are similar for scattering lengths of equal size but opposite
sign.

\section{ Repulsively bound pairs of fermions}

Although so far we have only discussed repulsively bound pairs which
are composed of {\em bosonic} atoms, analogous pairs can also be
formed from {\em fermions} or {\em boson-fermion mixtures}. These
systems will exhibit interesting physics based on their composite
nature and the quantum statistics of their components.

\begin{figure}[htbp]
\begin{center}
\includegraphics[width=0.25\textwidth]{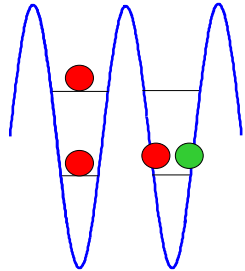}
\caption{Possible realization of repulsively bound pairs with
fermionic atoms. {\bf left} In the case of two identical fermions
(same spin) the atoms have to be in different bands due to the Pauli
exclusion principle.  {\bf right} Fermions of different spins or
Bose-Fermi mixtures, however, can occupy the same band.}
\label{Fig:fermions}
\end{center}
\end{figure}

Of course, in a single species fermion experiment it is not possible
to put two identical fermions into the same site and band due to the
Pauli exclusion principle. The two atoms would have to be at least
in different bands (see Fig. \ref{Fig:fermions}, left), and even
then the interaction between them typically would be very small in
the ultracold regime.

These problems do not arise using a two-component spin mix of fermions (see
Fig. \ref{Fig:fermions}, right), as two atoms of different spin can share the
same site and band and can also interact strongly. The fact that a higher site
occupancy than two is again strictly forbidden could be advantageous in the
initial production of pairs. Furthermore, the pairing of two fermions can
result in a pair with bosonic character. In a 3D environment pairing of
fermions recently lead to interesting experiments studying the BEC-BCS
transition (see e.g. \cite{Reg04,Chi04,Kin04b,Bou04,Zwi05,Par05}). It would be
interesting to study similar properties to this transition with repulsive
pairing, investigating the system as the interaction strength is changed.

Also it would be very interesting to study repulsive pairs which are a
composite objects of a fermion and a boson. One question would be how these
bound states, would interact with each other, and how, for example, the bosonic
atoms within the pair would mediate next neighbor interactions \cite{Lew04}.

\section{Other related physical systems}
Although no stable repulsively bound pairs have previously been
observed, their physics is partially related to other physical
systems. Here we briefly discuss a few examples of such systems with
bound states which are in a way reminiscent of repulsively bound
states.

\subsection{Pairing resonances in many-body systems}
For example, resonance behaviour based on similar pairing of
Fermions of different spin in the Hubbard model was first discussed
by Yang \cite{Yan1989}, and plays an important role in SO(5)
theories of superconductivity \cite{Han2004}. There are several
examples of many-body bound states that can occur for repulsive as
well as attractive interactions, such as the resonances discussed in
the context of the Hubbard model by Demler et al. \cite{Dem1995}.
Such resonance behaviour is common in many-body physics, although
states of this type are normally very short-lived. Optical lattice
experiments will now provide an opportunity to prepare and
investigate stable versions of such states, which until now have
only appeared virtually as part of complex processes.

\subsection{Excitons}
The stability and many-body physics of repulsively bound pairs is
perhaps most closely associated with that of excitons, which are
{\em attractively} bound pairs of a particle in the conduction band
and a hole in the valence band of a periodic system
\cite{Moskalenko}. These bind to form a composite boson, a gas of
which can, in principle, Bose-condense. Excitons are excited states
of the many-body system, but are bound by an attractive interaction
between the particle and hole that form the pair. They are also
discussed in the specific context of fermionic systems. However, a
single exciton on a lattice could have a description very similar to
that of a single repulsively bound pair, and could be realised and
probed in optical lattices experiments \cite{Kan06}.

\subsection{Photonic crystals}
Repulsively bound atom pairs in an optical lattice are also reminiscent of
photons being trapped by impurities in photonic crystals
\cite{photonicCrystal}, which consist of transparent material with periodically
changing index of refraction. An impurity in that crystal in the form of a
local region with a different index of refraction can then give rise to a
localized field eigenmode. In an analogous sense, each atom in a repulsively
bound pair could be seen as an impurity that ``traps'' the other atom.

\subsection{Gap solitons}
An analogy can also be drawn between repulsively bound atom pairs and gap
solitons, especially as found in atomic gases
\cite{Lou2003,Efr2003,Eie2004,Ahu04}. Solitons are normally a non-linear wave
phenomenon, and in this sense have a very different behaviour to repulsively
bound pairs, which exhibit properties characteristic of many-body quantum
systems. However, there has been increasing recent interest in discussing the
limit of solitons in atomic systems where very few atoms are present, giving
rise to objects that are often referred to as quantum solitons
\cite{quantumsolitons1,quantumsolitons2,quantumsolitons3}. These are N-body
bound states in 1D, and thus a 2-atom bright quantum soliton is a bound state
of two atoms moving in 1D. In this sense, the solution for a single repulsively
bound pair in 1D is related to a single quantum soliton on a lattice.

\section{Conclusion}
We have reviewed theoretically and experimentally the physics of repulsively
bound pairs of atoms in an optical lattice. The good agreement between
experiment and theory exemplifies the strong correspondence between the optical
lattice physics of ultracold atoms and the Hubbard model on a new level, a
connection which has particular importance for applications of these cold atom
systems to more general simulation of condensed matter models and to quantum
computing. The existence of such metastable bound objects will be ubiquitous in
cold atoms lattice physics, giving rise to new potential composite objects also
in fermions or in systems with mixed Bose-Fermi statistics. These states could
also be formed with more than two particles, or as bound states of existing
composite particles. Repulsively bound pairs have no direct counterpart in
condensed matter physics due to the strong inelastic decay channels observed in
solid state lattices, and could be a building block of yet unstudied quantum
many body states or phases.

\acknowledgments We would like to thank our fellow team members who
contributed to the research work on repulsively bound atom pairs:
Hanspeter B\"uchler, Rudi Grimm, Adrian Kantian, Florian Lang,
Gregor Thalhammer, Klaus Winkler, and Peter Zoller. We would like to
thank Eugene Demler for interesting discussions. We acknowledge
support from the Austrian Science Fund (FWF) within the
Spezialforschungsbereich 15, from the European Union within the
OLAQUI and SCALA networks, from the TMR network "Cold Molecules",
and the Tiroler Zukunftsstiftung.


\begin{thebibliography}{0}



\bibitem{hubbardtoolbox}   \BY{Jaksch~D. \atque  Zoller~P.}
\IN{Annals of Physics}{315}{2005}{52} \atque references therein.

\bibitem{Blo05}
\BY{ Bloch I.}
\IN{Nature Physics}{1}{2005}{23}.

\bibitem{Gre02}
\BY{ Greiner M., Mandel O., Esslinger T., H\"ansch T. W. \atque Bloch I.}
\IN{Nature}{415}{2002}{39}.

\bibitem{Par04}
\BY{Paredes B.,   Widera A., Murg V., Mandel O., F\"olling S., Cirac I.,
Shlyapnikov G. V.,  H\"ansch T.W., \atque Bloch I.}
 \IN{Nature}{429}{2004}{277}.

\bibitem{Sto04}
 \BY{St\"oferle T.,   Moritz H.,  Schori C.,  K\"ohl M. \atque   Esslinger T.}
\IN{Phys. Rev. Lett.}{92}{2004}{130403}.

\bibitem{Kin04}
\BY{Kinoshita T., Wenger T. \atque Weiss, D. S.}
\IN{Science}{305}{2004}{1125}.

\bibitem{Lab04}
 \BY{Laburthe Tolra B.,  O'Hara K. M.,  Huckans J. H.,  Phillips W. D.,  Rolston S. L.
  \atque  Porto J. V.}
\IN{Phys. Rev. Lett.}{92}{2004}{190401}.

\bibitem{Fal06}
 \BY{Fallani L.,  Lye J. E., Guarrera V.,  Fort C. \atque  Inguscio M.}
cond-mat/0603655 (2006).

\bibitem{Win2006}
\BY{Winkler~K., Thalhammer~G., Lang~F., Grimm~R., Hecker Denschlag ~J.,
Daley~A. J., A. Kantian~A., B\"uchler~H. P. \atque Zoller~P.}
 \IN{Nature}{441}{2006}{853}.

\bibitem{Hof02}
\BY{Hofstetter W., Cirac J. I., Zoller P., Demler E. \atque Lukin M. D.}
\IN{Phys. Rev. Lett.}{89}{2002}{220407}.

\bibitem{Lew04}
\BY{Lewenstein M., Santos L., Baranov M. A. \atque Fehrmann H.}
\IN{Phys. Rev. Lett.}{92}{2004}{050401}.

\bibitem{Fis89}
  \BY{Fisher~M. P. A.,  Weichman~P. B.,  Grinstein~G.,  \atque  Fisher~D. S.}
 \IN{Phys. Rev. B}{40}{1989}{546}.



\bibitem{alpharef}
 \BY{Cserti J.} \IN{American Journal of Physics}{68}{2000}{896}.

\bibitem{Fle06}
 \BY{Petrosyan D., Fleischhauer M., Anglin J. R.} cond-mat/0610198.

\bibitem{td-dmrg1}
\BY{Vidal~G.} \IN{Phys. Rev. Lett.}{91}{2003}{147902};
\SAME{93}{2004}{040502}.

\bibitem{td-dmrg2}
 \BY{Daley~A. J.,  Kollath~C.,  Schollw\"ock~U. \atque  Vidal~G.} \IN{J. Stat.
Mech.: Theory Exp.}  {P04005} {2004} {}.

\bibitem{td-dmrg3}
 \BY{White~S. R. \atque Feiguin~A. E.} \IN{Phys. Rev. Lett.}
{93}{2004}{076401}.

\bibitem{dmrgreview}
 \BY{Schollw\"{o}ck~U.} \IN{Rev. Mod. Phys.}{77} {2005}{259}.

\bibitem{vex1}
    \BY{Gobert D., Kollath C., Schollw\"ock U. \atque Sch\"utz G.}
    \IN{Phys. Rev. E}{71}{2005}{036102}.

\bibitem{vex2}
    \BY{Daley A. J., Clark S. R., Jaksch D. \atque Zoller P.}
    \IN{Phys. Rev. A}{72}{2005}{043618}.

\bibitem{vex3}
    \BY{Kollath C., Schollw\"ock U., \atque Zwerger W.} \IN{Phys. Rev. Lett.}{95}{2005}{176401}.

\bibitem{vex-spectroscopy}
    \BY{Kollath C., Iucci A., Giamarchi T., Hofstetter W. \atque Schollw\"ock U.} \IN{Phys. Rev.
    Lett.}{97}{2006}{050402}.

\bibitem{vex4}
    \BY{Clark S. R. \atque Jaksch D.} \IN{Phys. Rev. A}{70}{2004}{043612}.

\bibitem{vex5}
    \BY{Al-Hassanieh K. A., Feiguin A. E., Riera J. A. ,
    Busser C. A. \atque Dagotto E.} cond-mat/0601411.

\bibitem{vex-spectroscopy2}
    \BY{Kollath C., Iucci A., McCulloch I. \atque Giamarchi T.}
    cond-mat/0608091.

\bibitem{Vid04}
    \BY{Zwolak M. \atque Vidal G} \IN{Phys. Rev. Lett.}{93}{2004}{207205}.

\bibitem{Cir04}
  \BY{Verstraete~F.,  Garcia-Ripoll~J. J. \atque
 Cirac J. I.} \IN{Phys. Rev. Lett.} {93}{2004}{207204}.

\bibitem{trotter1}
 \BY{Suzuki~M.} \IN{Phys. Lett. A}{146}{1990}{6}.

 \bibitem{trotter2}
 \BY{Suzuki~M.} \IN{J. Math. Phys.}{32}{1991}{2}.


\bibitem{Tha05} \BY{
Thalhammer~G., Theis~M., Winkler~K.,
 Grimm~R. \atque  Hecker Denschlag~J.}
\IN{Phys. Rev. A}{71}{2005}{033403}.

\bibitem{Gre01}  \BY{Greiner M.,  Bloch I.,   H\"ansch T. W., \atque
  Esslinger T.}
\IN{Phys. Rev. A}{63}{2001}{031401(R)}.


\bibitem{QUIC}
  \BY{ Esslinger T.,  Bloch I., \atque  H\"ansch T. W.}
\IN{Phys. Rev. A}{58}{1998}{2664(R)}.

\bibitem{Tha06} \BY{Thalhammer~G.,  Winkler~K.,  Lang~F.,  Schmid~S.,
 Grimm~R. \atque  Hecker
Denschlag~J.} \IN{Phys. Rev. Lett.}{96}{2006}{050402}.

\bibitem{Volz03}
 \BY{Volz T., D\"urr S., Ernst S., Marte A. \atque  Rempe, G.}
\IN{Phys. Rev. A}{68}{2003}{010702}.


\bibitem{Den02}
\BY{Hecker Denschlag J.,
  Simsarian J. E.,  H\"affner H.,  McKenzie C.,  Browaeys A.,  Cho D.,
 Helmerson K.,  Rolston S. L., \atque  Phillips W. D.}
\IN{J. Phys. B} {35}{2002}{3095}.

\bibitem{Reg04}
\BY{ Regal C. A.,  Greiner M. \atque  Jin D. S.} \IN{Phys. Rev.
Lett.}{92}{2004}{040403}.

\bibitem{Chi04}
\BY{ Chin C.,  Bartenstein M.,  Altmeyer A.,  Riedl S.,  Jochim S.,
 Hecker Denschlag J. \atque  Grimm R.}
\IN{Science}{305}{2004}{1128}.

\bibitem{Kin04b}
\BY{ Kinast J.,  Hemmer S. L.,  Gehm M. E.,  Turlapov A., \atque Thomas J. E.}
\IN{Phys. Rev. Lett.}{92}{2004}{150402}.

\bibitem{Bou04} \BY{  Bourdel T.,  Khaykovich L.,  Cubizolles J.,
 Zhang J.,  Chevy F.,  Teichmann M.,  Tarruell L.,
Kokkelmans S. J. J. M. F.  \atque C. Salomon}
 \IN{Phys. Rev. Lett.}{93}{2004}{050401}.

\bibitem{Zwi05}
\BY{ Zwierlein M. W.,  Abo-Shaeer J. R.,  Schirotzek A.,  Schunck C. H., \atque
Ketterle W.} \IN{Nature}{435}{2005}{1047}.

\bibitem{Par05}
\BY{  Partridge G. B.,  Strecker K. E.,  Kamar R. I.,  Jack M. W., \atque Hulet
R. G.} \IN{Phys. Rev. Lett.}{95}{2005}{020404}.



\bibitem{Yan1989}
 \BY{Yang~C.N.} \IN{Phys. Rev. Lett.}{63}{1989}{2144}.

\bibitem{Han2004}
 \BY{Demler~E.,  Hanke~W. \atque  Zhang~S. C.} \IN{Rev. Mod. Phys.}
{76}{2004}{909}.

\bibitem{Dem1995}
 \BY{Demler~E. \atque  Zhang~S. C.} \IN{Phys. Rev. Lett.}{75}{1995}
{4126}.


\bibitem{Moskalenko}
 \BY{Moskalenko~S. A. \atque  Snoke~D. W.} \TITLE{Bose-Einstein
Condensation of Excitons and Biexcitons} (Cambridge University
Press, Cambridge, 2000).

\bibitem{Kan06} \BY{Kantian~A. et al.} in preparation.

\bibitem{photonicCrystal}
  \BY{Joannopoulos~J. D.,   Meade~R. D. \atque   Winn~J. N.}
\TITLE{Photonic Crystals: Molding the Flow of Light}, Princeton
University Press, Princeton, 1995.


\bibitem{Lou2003}
 \BY{Louis~P. J. Y.,  Ostrovskaya~E. A.,  Savage~C. M. \atque  Kivshar
~Yu. S.} \IN{Phys. Rev. A}{67}{2003}{013602}.

\bibitem{Efr2003}
 \BY{Efremidis~N. K. \atque  Christodoulides~D. N.}
 \IN{Phys. Rev. A}{67}{2003}{063608}.

\bibitem{Eie2004}
 \BY{Eiermann~B.,  Anker~Th.,  Albiez~M.,  Taglieber~M., Treutlein~P.,
Marzlin~K. P. \atque  Oberthaler~M. K.} \IN{Phys. Rev.
Lett.}{92}{2004}{230401}.

\bibitem{Ahu04}\BY{Ahufinger~V.,  Sanpera~A.,  Pedri~P.,  Santos~L. \atque
Lewenstein~M.} \IN{Phys. Rev. A}{69}{2004}{053604}.

\bibitem{quantumsolitons1}
 \BY{Drummond~P. D.,  Kheruntsyan~K. V. \atque  He~H.} \IN{J. Opt. B:
Quant. Semiclass. Optics}{1}{1999}{387}.

\bibitem{quantumsolitons2}
 \BY{Bullough~R. K. \atque  Wadati~M.} \IN{J. Opt. B: Quant. Semiclass.
Optics}{6}{2004}{S205}.

\bibitem{quantumsolitons3}
\BY{Mazets~I. E. \atque  Kurizki~G.} \IN{Eur. Phys.
Lett.}{76}{2006}{196}.

\end{thebibliography}
\end{document}